\documentclass[11pt]{article}
\usepackage{moriond}
\pdfoutput=1

\providecommand{\pt}{\ensuremath{p_\mathrm{T}}}
\providecommand{\kt}{\ensuremath{k_\mathrm{T}}}
\providecommand{\alps}{\ensuremath{\alpha_S}}
\providecommand{\alpsq}{\ensuremath{\alpha_S(Q)}}
\providecommand{\alpsmz}{\ensuremath{\alpha_S(M_Z)}}
\providecommand{\avept}{\ensuremath{\langle p_\mathrm{T1,2}\rangle}}
\providecommand{\Ratio}{$R_\mathrm{32}$}

\newcommand{\Photo}{\includegraphics[height=35mm]{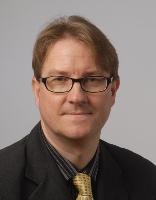}}

\begin{document}
\vspace*{4cm} \title{LATEST JETS RESULTS FROM THE LHC}

\author{K. RABBERTZ\\(on behalf of the ATLAS and CMS Collaborations)}

\address{Karlsruher Institut f{\"u}r Technologie,
  Institut f{\"u}r Experimentelle Kernphysik,\\
  Campus S{\"u}d, Postfach 6980, D-76128 Karlsruhe, Germany}

\maketitle\abstracts{%
  A fundamental characteristic of hadron colliders is the abundant
  production of jets, which then are studied to learn about hard QCD,
  the proton structure, or nonperturbative effects. In the following
  the latest results and developments from the LHC experiments on jet
  cross sections, event shapes, flavour and rapidity dependence, and
  cross section ratios are presented. The ratio of the inclusive 3-jet
  to the inclusive 2-jet event cross section is used for a first
  determination of the strong coupling constant at the TeV scale.%
}

\section{Jet Cross Sections}
\label{sec:crosssections}

Yields of collimated streams of particles, i.e.\ jets, are among the
most fundamental observables at hadron colliders. Through their high
production rates at low jet transverse momenta, \pt, they serve to
benchmark detector performance and to perform first cross section
measurements that are compared to theory predictions. As more data are
accumulated, ever higher jet \pt's become accessible providing
significant constraints on the parton distribution functions (PDFs) of
the proton and important input to searches for new physics.

ATLAS and CMS,\cite{Aad:2008zzm,Chatrchyan:2008aa} both employ the anti-\kt\ jet
algorithm\cite{Cacciari:2008gp} to define their jets,
but with different jet size parameters $R$ of $0.4$ or $0.6$ for ATLAS
and $0.5$ or $0.7$ for CMS respectively. The dominant source of
experimental uncertainty for jet measurements is the jet energy
calibration, because the steeply falling jet \pt~spectrum induces an
approximately $5$--$6$ times larger uncertainty on the jet cross
sections than on the jet energies. Profiting from the excellent
performance of both detectors, jet energy calibration uncertainties
could be limited to about $1$--$3$\% already during the initial
three-years running period, a feat that could be achieved previously
only after many years. Apart from jets with less than $\approx 50$ GeV
of transverse momentum this good performance could be kept up despite
the deteriorating influence of more pile-up collisions, i.e.\
additional proton-proton collisions in the same or neighbouring bunch
crossings. At the next LHC start-up with even higher instantaneous
luminosities pile-up will again pose a challenge.

The common normalization uncertainty caused by the luminosity
determination could be reduced from initially more than $10$\% down to
$2$--$4$\%. Including other systematic effects the inclusive jet
\pt~or the dijet mass cross sections could be measured by ATLAS and
CMS to roughly $10$ to $20$\% accuracy with less precision at the low
end and also less data at the high end of the jet \pt~and dijet mass
ranges.\cite{Aad:2011fc,ATLAS:2012jpa,Chatrchyan:2012bja}

It is a great achievement that these data are in general agreement
with predictions of QCD over many orders of magnitude in cross
section. With theoretical uncertainties of a similar order or larger
these jet measurements help constrain PDFs. Recent progress in
theory towards next-to-next-to-leading order (NNLO) predictions are
reported elsewhere in these proceedings.\cite{Pires:2013Moriond}

\section{Event Shapes}
\label{sec:shapes}

Alternatively to jet counting continuous dimensionless quantities that
are not sensitive to the details of soft nonperturbative effects of
QCD can be defined to characterize events. They are called {\em event
  shapes}\/ and have been in use since the early days of QCD in the
70s. Several collinear and infrared-safe event shapes, e.g.\ {\em
  transverse thrust}, have been investigated by ATLAS and CMS, where
the emphasis was on hard proton-proton scatterings and the phase space
is subdivided into bins of \pt~of the leading jet or the sum of all
jet \pt's.\cite{Aad:2012np,Khachatryan:2011dx} In a new study ATLAS
uses charged particles measured in their tracking system to
investigate event shapes down to very small transverse momenta of
$0.5$ GeV\@.\cite{Aad:2012fza} In accordance with previous results it
is found that better tunings of the MC event generators are needed in
order to describe all available data.

\section{Flavour and Rapidity Results}
\label{sec:rapflavour}

\begin{figure}
  \centering
  \includegraphics[width=0.33\linewidth]{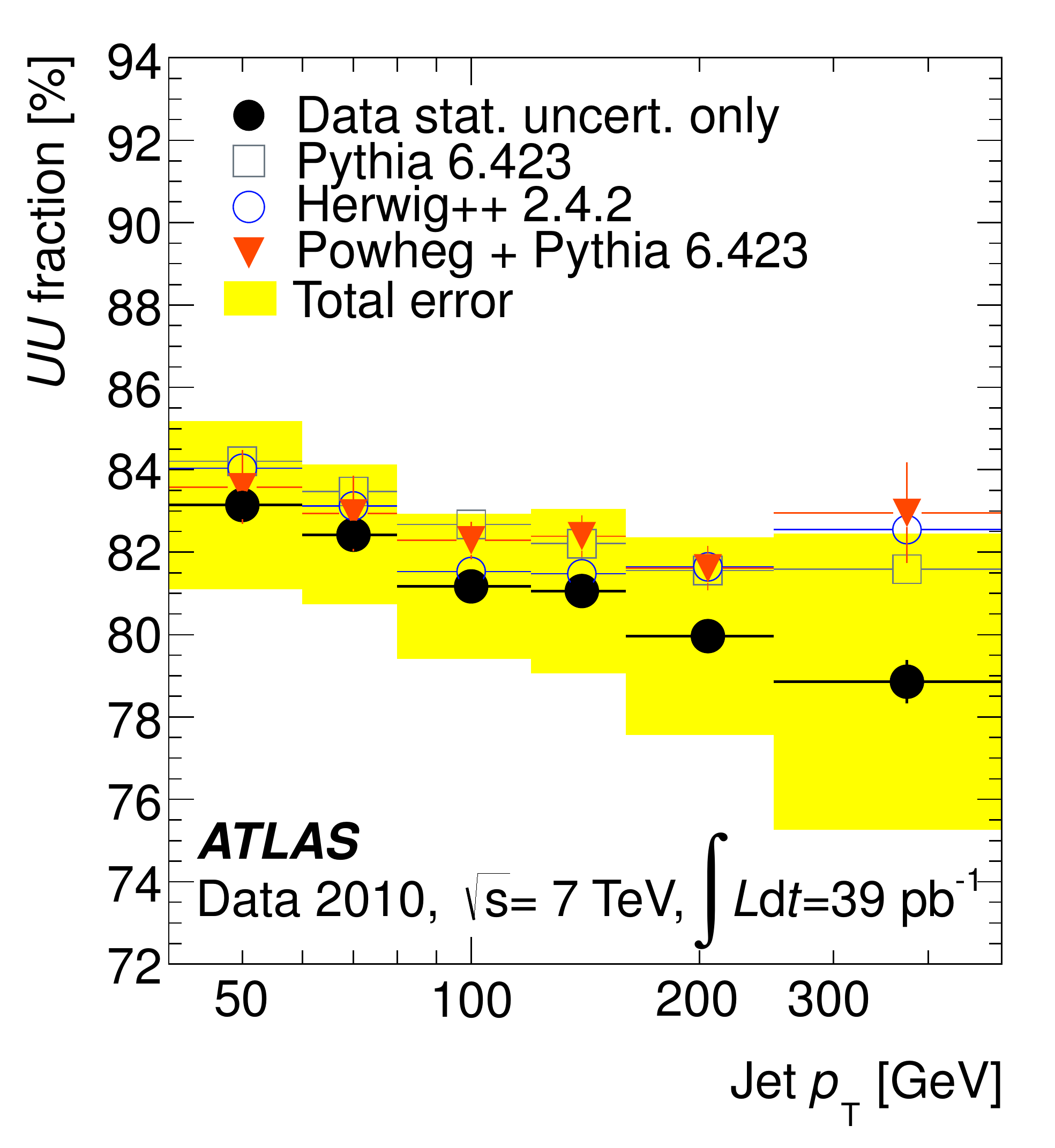}\hfill%
  \includegraphics[width=0.33\linewidth]{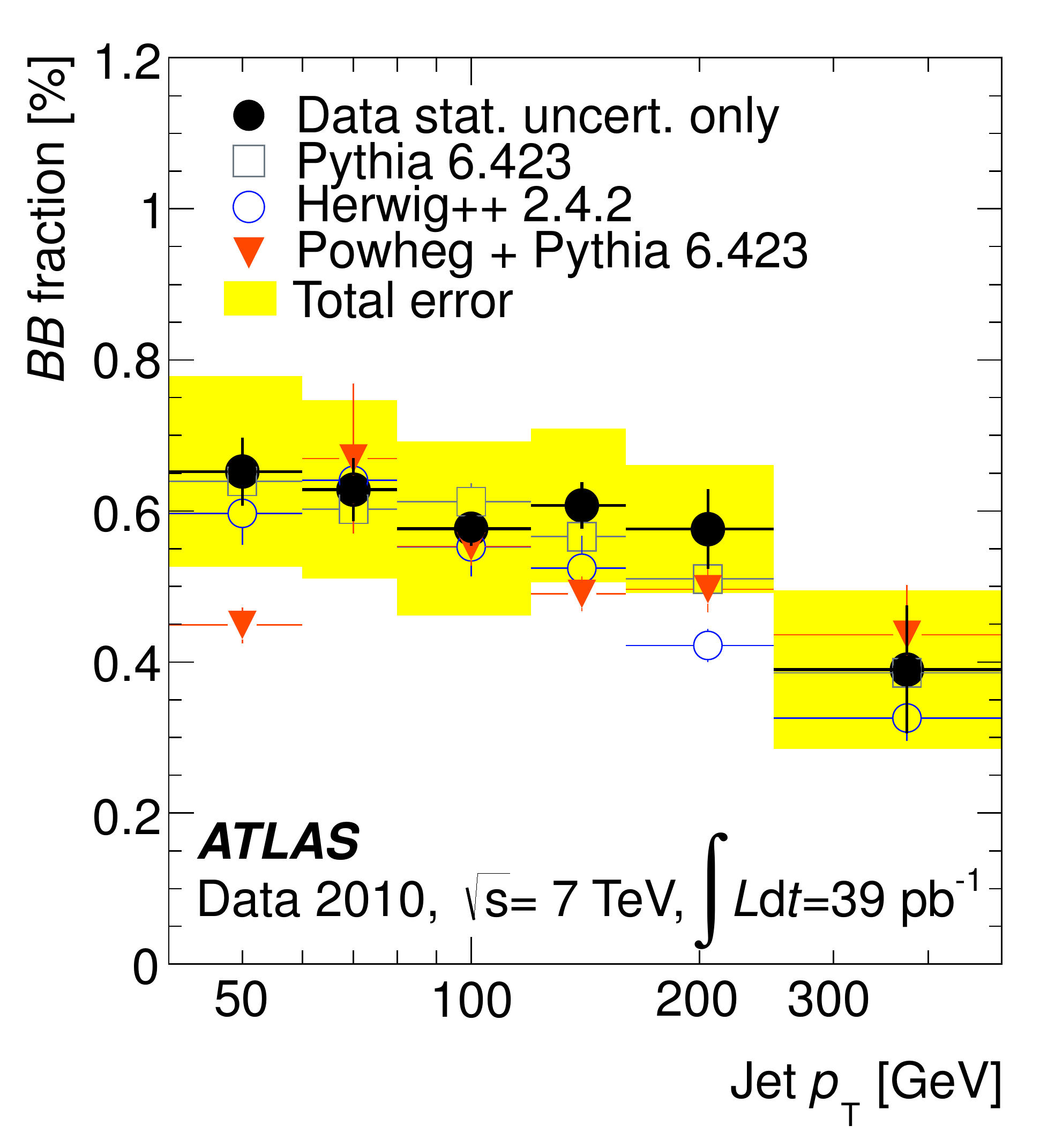}\hfill%
  \includegraphics[width=0.33\linewidth]{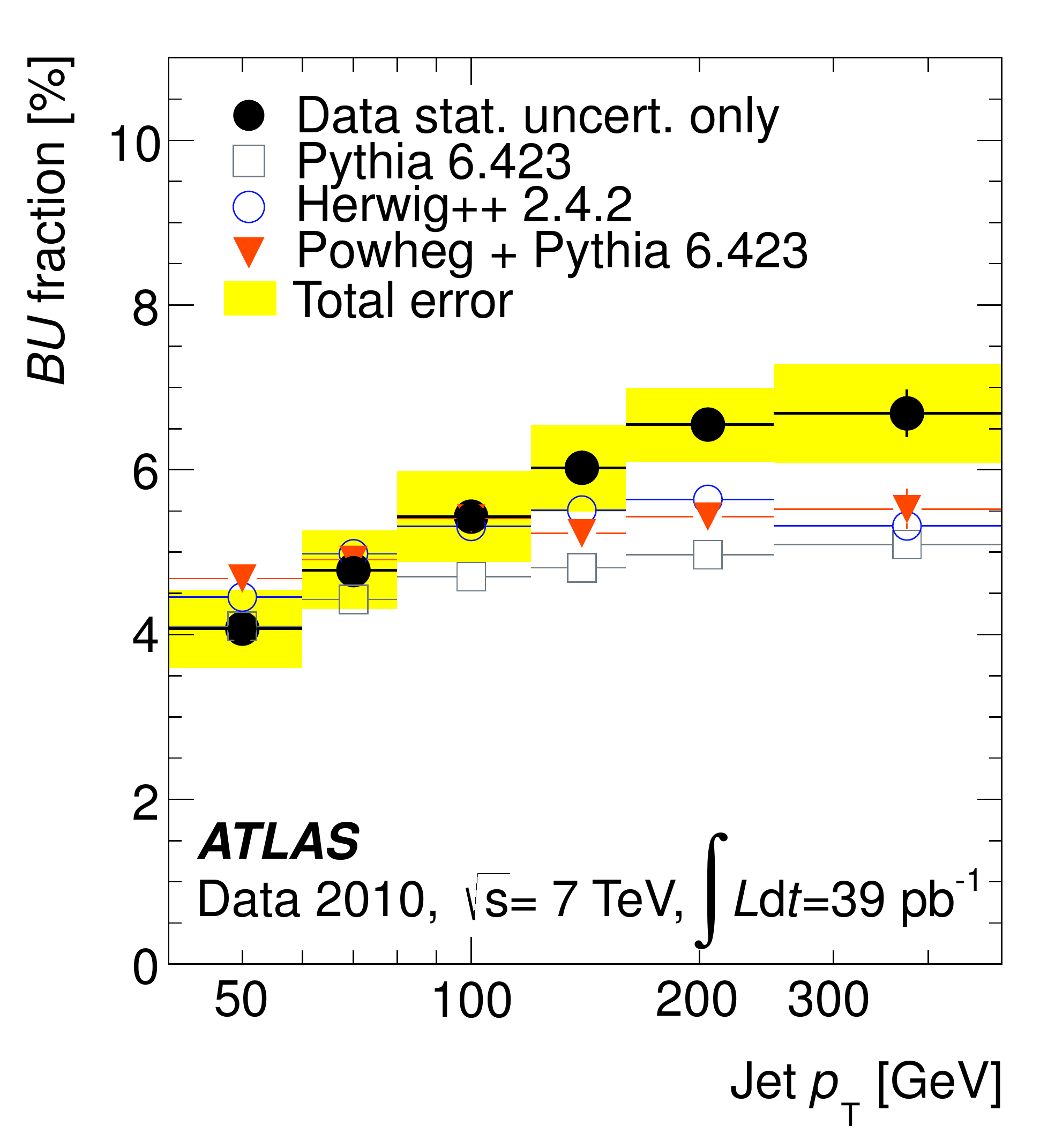}
  \caption{Relative frequency of jet flavours $B$, $C$, and light
    ($U$) in dijet events for the combinations $BB$ (left), $UU$
    (middle), and $BU$ (right plot).}
  \label{fig:dijetflavours}
\end{figure}

In Ref.~\cite{Aad:2012ma} ATLAS compares the relative frequency of jet
flavours in dijet events. Using template fits to kinematic properties
of secondary vertices inside jets, ATLAS differentiates between the
heavy, $b$ and $c$ quark initiated jets and {\em light}\/ jets, which
are labelled as $B$, $C$, and $U$
respectively. Figure~\ref{fig:dijetflavours} shows exemplarily the
measurements for the $UU$, $BB$, and $BU$ dijet combinations, which
demonstrate that light dijets compose about $80$\% of the total cross
section while the heavy $BB$ combination appears only in half a
percent of the events. The results are described well within
uncertainties by MC predictions except for the $BU$ case, where some
discrepancies can be observed.

Other analyses by ATLAS and CMS look into the dependence of the most
forward and backward dijets or the ratio of all possible dijet pair
distances over the leading dijet pair distance versus their separation
in absolute rapidity, $|\Delta y|$.\cite{Aad:2011jz,Chatrchyan:2012pb}
It is expected that deviations from the usual evolution of the PDFs
could show up in these quantities, where small parton momentum
fractions are accessed. The comparisons of different models to the
data are, however, inconclusive so far.

\section{Ratios of Jet Cross Sections and Determination of the Strong
  Coupling Constant}
\label{sec:ratios}

In order to reduce experimental as well as theoretical uncertainties,
jet cross section ratios are considered. Disposing of data at $2.76$
TeV, the baseline proton-proton centre-of-mass energy for heavy ion
collisions, in addition to the $7$ TeV data, ATLAS analyzed the ratio
of the inclusive jet cross section at these two energy
points.\cite{Aad:2013lpa} This provides more significant constraints
on PDFs in the accessible phase space than considering each jet cross
section separately.

The ALICE Collaboration followed a suggestion in
Ref.~\cite{Soyez:2011np} and examined the cross section ratio for jets
defined with different jet size parameters, which in their case are
$R=0.2$ and $0.4$ because of their primary focus on heavy ion
physics.\cite{Abelev:2013fn} Using this method, details of the parton
showering and the nonperturbative hadronization phase are emphasized
in this ratio such that even NLO calculations are not able to describe
the data as shown by ALICE\@.\cite{Aamodt:2008zz}

\begin{figure}[t]
  \centering
  \includegraphics[width=0.46\linewidth]{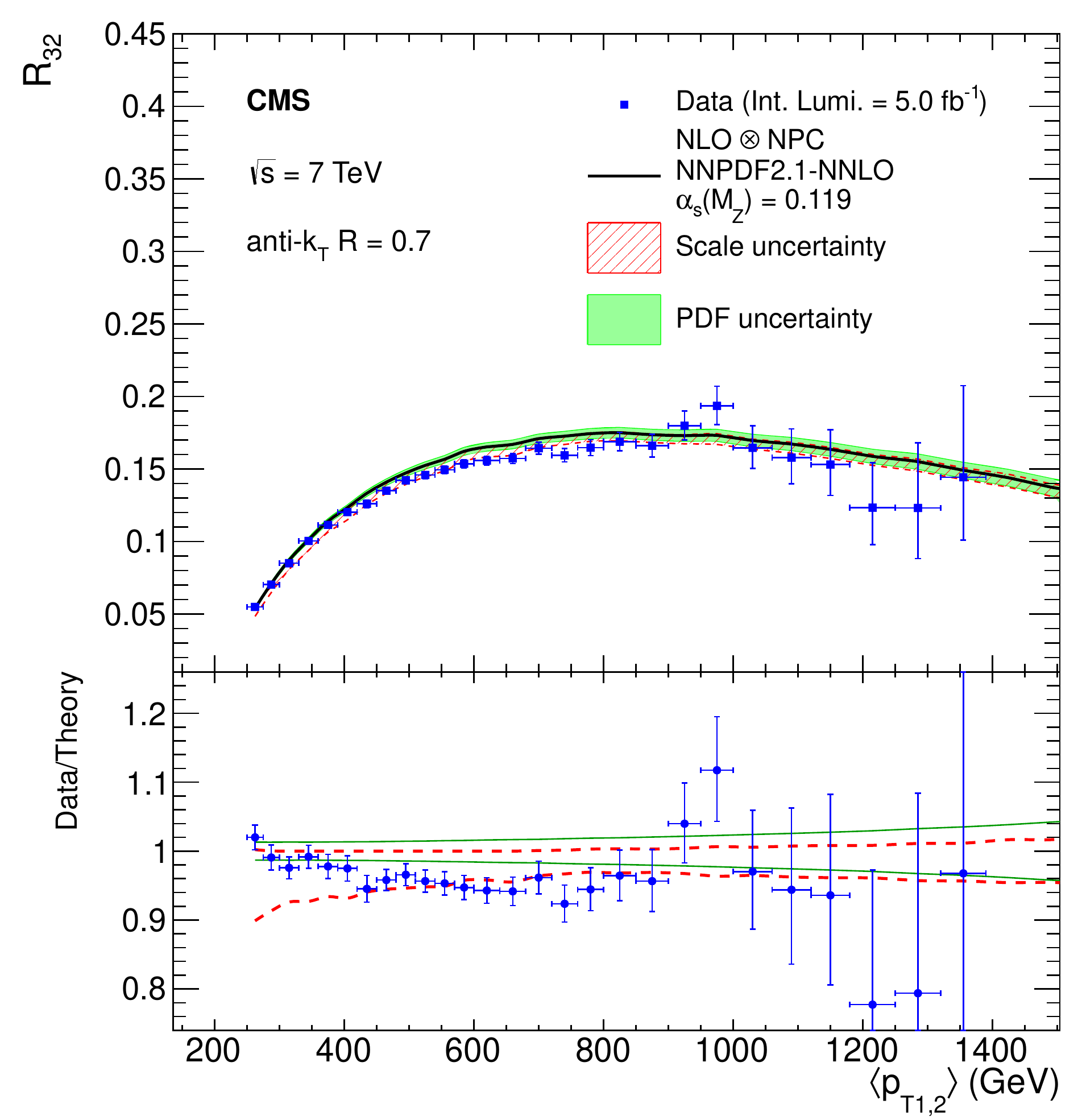}\hfill%
  \includegraphics[width=0.5\linewidth]{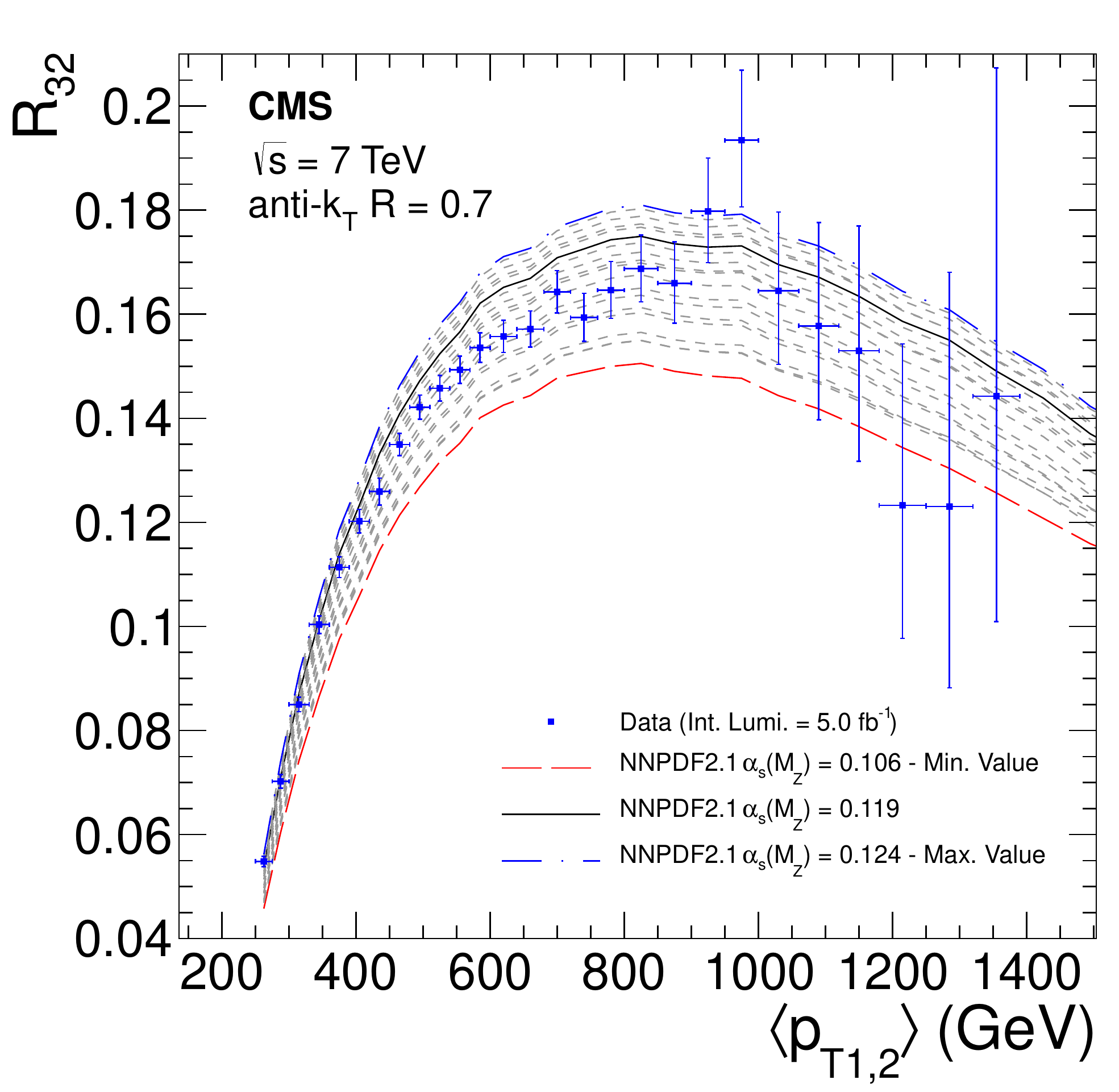}
  \caption{Left: Measured ratio \Ratio\ versus \avept\ (solid circles)
    together with
    the NLO prediction (solid line) corrected for nonperturbative
    effects (NPC), the scale uncertainty, and the PDF uncertainty.
    The bottom panel shows the ratio of data to the theoretical
    predictions, together with bands representing the scale (dotted
    lines) and PDF (solid lines) uncertainties while the error bars
    correspond to the total uncertainty. Right: The same comparison
    with theory but for a series of values of \alpsmz. The \alpsmz\
    value is varied in the range of $0.106$–-$0.124$ in steps of
    0.001. In both cases the NNPDF2.1 PDF set at NNLO evolution order
    has been employed.}
  \label{fig:assensitivity}
\end{figure}

\begin{figure}[t]
  \centering
  \includegraphics[width=0.70\linewidth]{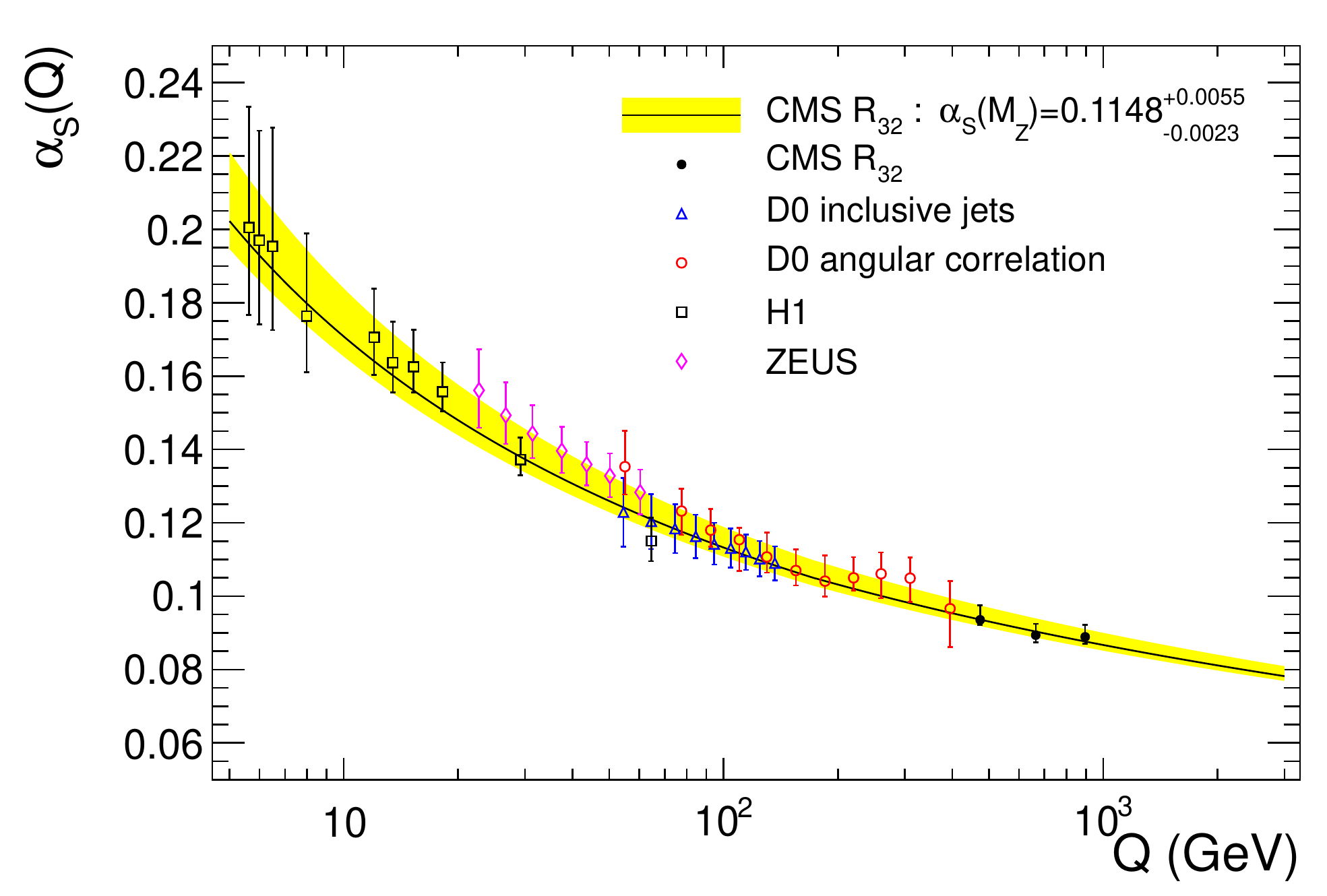}
  \caption{The strong coupling \alpsq\ (solid line) and its total
    uncertainty (band) evolved from the CMS determination
    $\alpsmz=0.1148_{-0.0023}^{+0.0055}$ as a function of the momentum
    transfer $Q=\avept$. The extractions of \alpsq\ in three separate
    ranges of $Q$ are shown together with results from other hadron
    collider experiments.}
  \label{fig:asrunning}
\end{figure}

In contrast, the ratio of the inclusive 3-jet to the inclusive 2-jet
event cross section, \Ratio, has been demonstrated by CMS to be
reliably comparable to perturbative QCD, if sufficiently high
thresholds, CMS requires $p_\mathrm{T,jet} > 150$ GeV, are imposed on
all jets including the third leading one.\cite{Chatrchyan:2013txa}
Figure~\ref{fig:assensitivity} presents a comparison of the data,
differential in the average transverse momentum of the two leading
jets, \avept, to the theory predictions at NLO on the left and the
sensitivity to a variation of the strong coupling constant \alpsmz\ on
the right for the NNPDF2.1 PDF set at NNLO evolution
order.\cite{Ball:2011mu} Fits of \Ratio\ in the range of $420 < \avept
< 1390$ GeV have been used to determine the strong coupling constant
\alps\ at the scale of the Z boson mass to be:
\begin{equation}
  \label{eqn:total_result}
  \alpsmz = 0.1148 \pm 0.0014\,\mathrm{(exp.)}  \pm
  0.0018\,\mathrm{(PDF)} ^{+0.0050}_{-0.0000}\,\mathrm{(scale)} =
  0.1148\,_{-0.0023}^{+0.0055},
\end{equation}
compatible with the world average value of
$\alpsmz=0.1184\pm0.0007$.\cite{Beringer:1900zz} Here, the total
uncertainty is derived from the experimental, PDF, and scale
uncertainties by quadratic addition. This is the first determination
of the strong coupling constant from measurements at scales $Q$ of the
order of 1 TeV\@. A comparison to other determinations at hadron
colliders accessing different scales $Q$ is shown in
Fig.~\ref{fig:asrunning}. No deviation from the expected running
behaviour of the strong coupling constant is observed.

\section{Summary}
\label{sec:summary}

In summary, hadron colliders, which are usually conceived of as {\em
  discovery machines}, are also great {\em jet laboratories}, which
provide many opportunities for precise measurements as demonstrated.
Further results from the Tevatron are discussed
elsewhere.\cite{Wobisch:2013Moriond} Through these measurements more
insight can be gained into the workings of QCD, our theory of the
strong interaction, with significant impact on e.g.\ other cross
section predictions and searches for new physics.


\section*{References}
\bibliography{moriond}

\end{document}